\documentclass[a4paper]{jacow}

\ifboolexpr{bool{xetex} or bool{luatex}} 
 {}                                      
\usepackage[utf8]{inputenc}           

\usepackage[USenglish]{babel}			 

\usepackage[final]{pdfpages}
\usepackage{multirow}
\usepackage{ragged2e}
	
\begin{document}

\title{Concept Study of a Compact, high gradient X-band Travelling wave RF photogun with Novel Laser Coupling Scheme}

\author{T. G. Lucas \thanks{Email: t.g.lucas@tue.nl}, P. H. A. Mutsaers, O. J. Luiten\\ Eindhoven University of Technology, The Netherlands.}
\maketitle

\begin{abstract}
This paper presents the design of a travelling wave RF photogun operating at the European X-band frequency of 11.994 GHz. The design is based on a CLIC prototype structure milled from copper halves and uses the unique gap introduced in the geometry to couple in the laser without obstructing the outgoing electron beam. With a modest peak input power of 19 MW and a fill time of 48 ns, the gun has the ability to operate at very high RF pulse repetition rates in comparison to standing wave RF photoguns. For its nominal input power, the cathode has a peak gradient of 91 MV/m and the entire gun has a gradient of 65 MV/m. The gun is illustrated to provide an 80~pC bunch at a 14.15~MeV with a 0.14~$\%$ RMS energy spread and an emittance of 0.6~mm~mrad at a repetition rate of at least 450~Hz.

\end{abstract}

\section{Introduction}

Future compact light sources are looking towards high gradient linear accelerators (linacs) to reduce their overall footprint as well as improve beam quality. Such linear accelerators have been the focus of the CLIC programme, looking at a future linear collider, which has illustrated that X-band linacs can operate at gradients in excess of 100 MV/m~\cite{J.Navarro16,R.Zennaro17,T.Lucas18}. For this reason, several compact X-ray source projects have focused on using X-band technology in their design~\cite{GravesICS,smartlight,compactlight}. Injecting into such facilities is typically addressed through a standing-wave (SW) RF photogun or using a DC gun with a bunching section. Each of these are robust methods but also have their drawbacks. For example, SW RF photoguns have long fill times and consequently are limited in their repetition rate. On the other hand, DC guns can pulse continously, or even produce DC beam, but their low cathode fields, in comparison to RF photoguns, limit the bunch charge possible for a given emittance. A more recent concept is the development of the travelling wave RF photogun which has become more feasible through the development of high gradient linacs~\cite{PSI_TWgun2014,PSI_TWgun2016,PSI_TWgun2017}. In~\cite{PSI_TWgun2017}, such a gun could be developed from an existing accelerating structure design through two modifications: 
\begin{enumerate}
    \item Adjust the length of the first cells to allow low energy capture, and
    \item Modify the input coupler to house a cathode.
\end{enumerate}
This paper details the design of the first x-band travelling wave RF photogun developed from a prototype high gradient accelerating structure designed for the CLIC programme, the CLIC-G Open~\cite{halves2018}. The gun will use the unique geometry of this prototype to couple in the laser through a gap incorporated into the cells' design. The paper will begin with a brief review of the RF design of the original accelerating structure followed by a description of the two significant modifications made to the CLIC-G Open to transform it into a travelling wave RF photogun. Following will be a description of the unique laser coupling scheme illustrating the benefits and consequences of the method. The RF fields from the gun design will be incorporated into a particle tracing software where a main solenoid and bucking coil will be introduced to combat space-charge effects in the first section of the gun. Finally, the paper will conclude by discussing the beam quality produced by the gun.

\begin{figure}
    \centering
    \includegraphics[width=0.8 \linewidth]{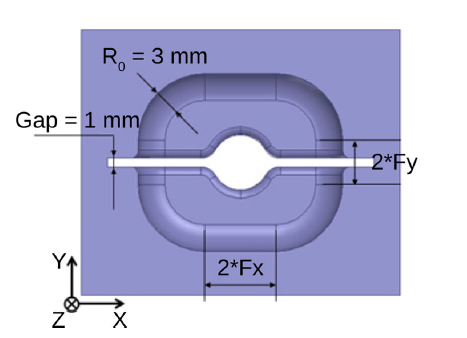}
    \caption{Cross-section of the cell geometry used for the CLIC-G Open structure with its "race-track" design and gap to house the joint~\cite{halves2018}.}
    \label{fig:celldesignoriginal_CS}
\end{figure}

\begin{figure}
    \centering
    \includegraphics[width= \linewidth]{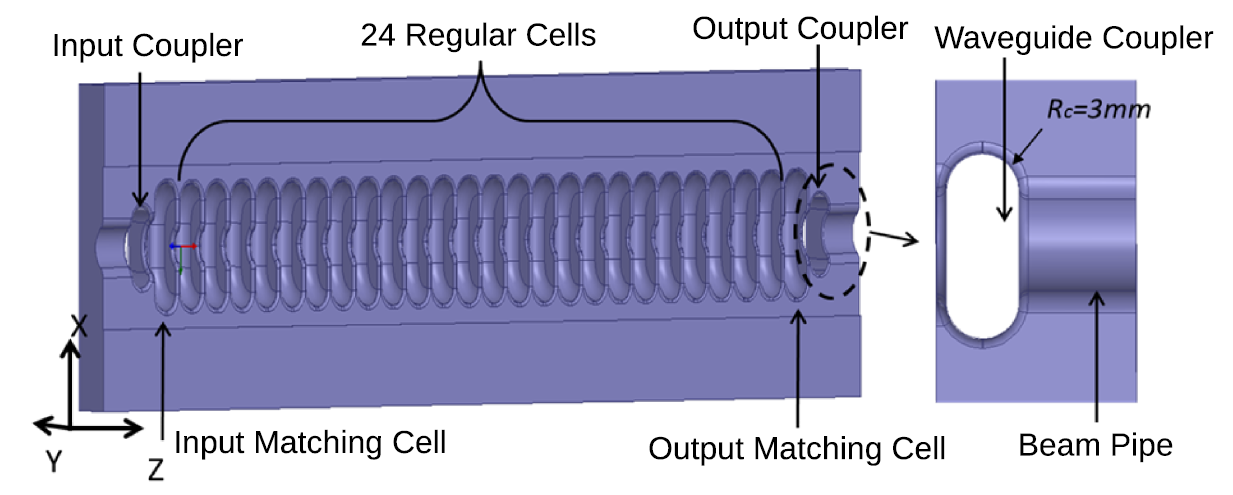}
    \caption{One half of the completed 24 cell structure of the CLIC-G Open structure~\cite{halves2018}.}
    \label{fig:halves_structure}
\end{figure}

\begin{figure*}
    \centering
    \includegraphics[width=0.8\linewidth]{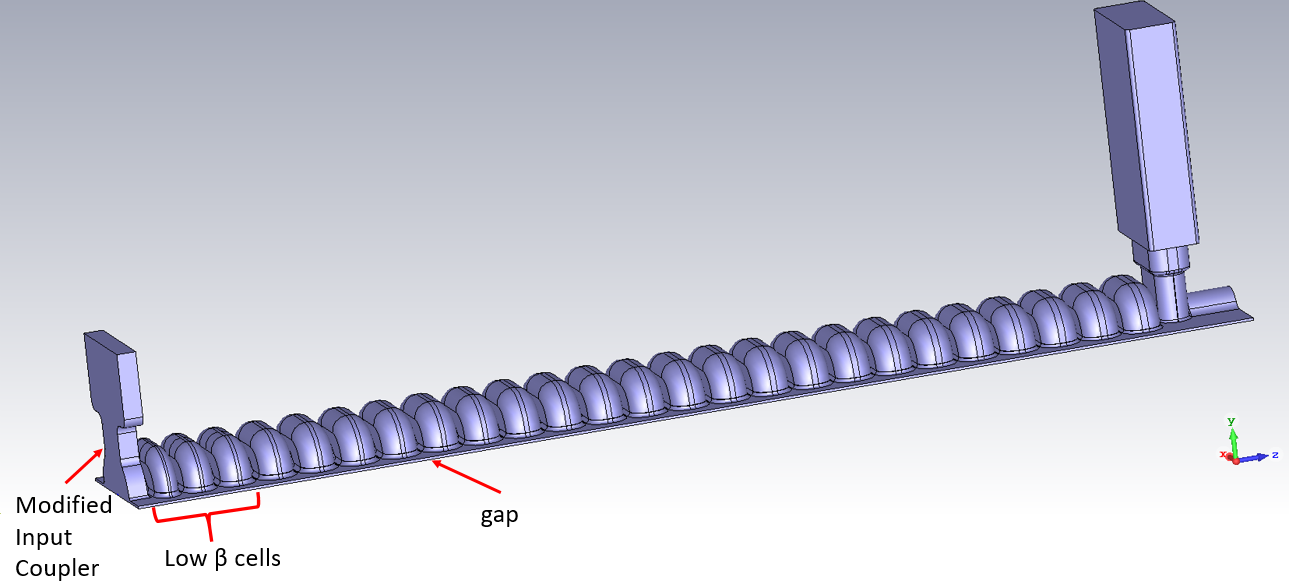}
    \caption{One symmetric half of the RF design (vacuum) of the travelling wave RF photogun with the notable features labelled.}
    \label{fig:structuredesign}
\end{figure*}

\section{RF Design}

\begin{figure*}
    \centering
    \includegraphics[width= \linewidth]{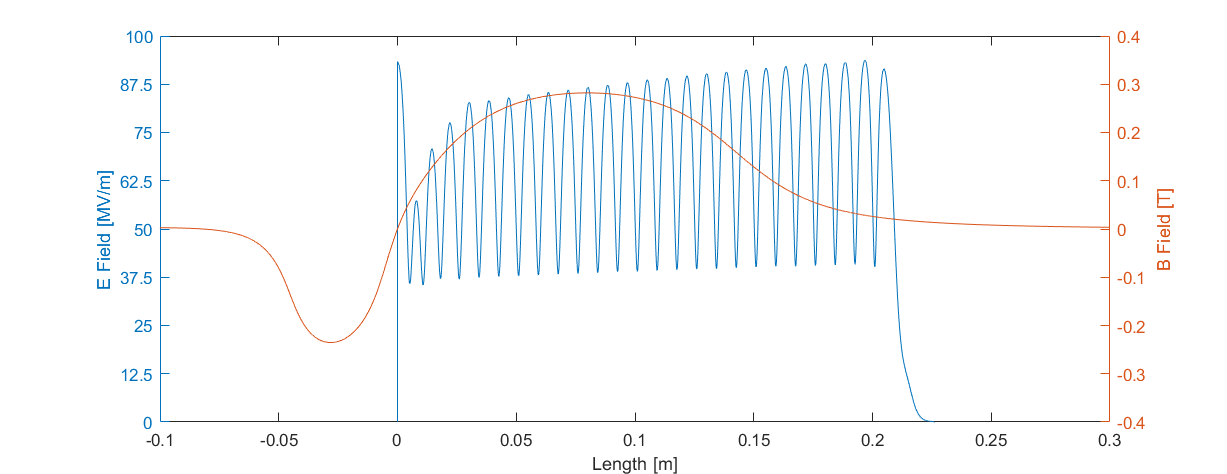}
    \caption{The axial longitudinal field distribution for the electric field from the accelerating structure and the magnetic field from the main solenoid and bucking coil.}
    \label{fig:EandHfield}
\end{figure*}

\begin{figure}
    \centering
    \includegraphics[width=\linewidth]{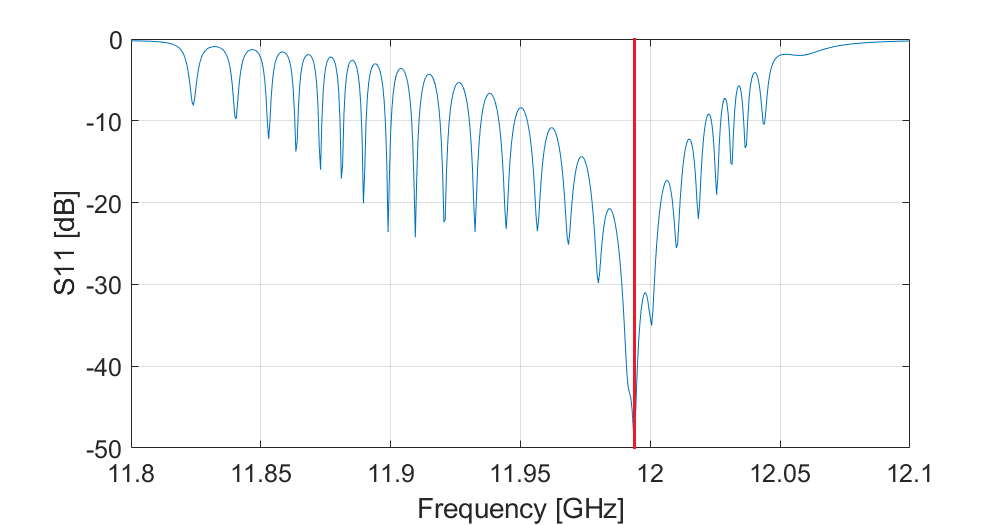}
    \caption{S11 parameter of the RF photogun.}
    \label{fig:S11}
\end{figure}

The RF design started with the CLIC-G Open structure which has been well-documented in~\cite{halves2018} therefore only an brief description will be discussed below. The design started with the CLIC-G prototype, the most widely tested CLIC prototype, which operates at a 120 degrees phase advance and the European X-band frequency of 11.994 GHz. With the aim of joining the cells through the milling of two halves, a gap was introduced into the design to house the brazing joint (Figure~\ref{fig:celldesignoriginal_CS}).
\\
The gap's width was crucial to cutting off the fundamental mode frequency preventing the fields from propagating into the gap and reaching the joint. This allows a variety of joining techniques to be used. The cells were designed with a "race-track" geometry where the cross-sectional profile includes straight regions of half-length Fx and Fy (Figure~\ref{fig:celldesignoriginal_CS}). These cells were ultimately arranged into a 24 cell structure, completed with a waveguide coupler at each end. With this as the initial geometry, the design of the travelling wave RF photogun could begin (Figure~\ref{fig:halves_structure}).
\\
The first task was to reduce the length of the first three regular cells, noting that the first cell in the structure is the input coupler cell where the cathode will later be housed. An analysis performed by X. Stragier illustrated that the lengths 6.832mm, 7.332 mm and 7.832~mm for the first, second and third cells, respectively, were an appropriate reduction in the cells' lengths for capture of the low beta electrons~\cite{tglucasHG2019}. The lengths of these cells were adjusted in the CLIC-G Open and then tuned back to the correct operational frequency. Following, the input coupler was to be modified to house a cathode. To do so, the input coupler was completely removed and replaced with a compact coupler design used in CLIC G* Structures~\cite{CLICGStar}. Additionally, a 1~mm gap was added on the plane normal to the RF input, as per the original CLIC-G Open coupler design, and the beampipe was replaced with a flat plane where the cathode would be placed. This RF design does not look at how the cathode would be installed although this has been addressed in~\cite{PSI_TWgun2017}. The input coupler's dimensions were then optimised until the reflection was below -45 dB. Various input coupler cell lengths were tested using General Particle Tracer (GPT). A length of 5~mm produced the lowest emittance for an input power of 19 MW. The input coupler was not optimised to reduce surface fields. The electric field along the z-axis of the final design is illustrated in Figure~\ref{fig:EandHfield}. The electric field at the cathode (length = 0~m) is observed to be 91~MV/m. Along with this, the first three cells are observed to have a reduced axial field strength. The RF frequency spectrum of the accelerator is plotted in Figure~\ref{fig:S11}, with the operational frequency marked with a red line, and Table~\ref{tab:RFparameters} lists the important RF parameters. The fill time is similar to that in the CLIC-G Open as expected. Given that the fill time is only 48 ns (Table~\ref{tab:RFparameters}), the design opens up the possibility to operate at very high repetition rates well above those possible in a current state-of-the-art SW RF photoguns. The CLIC-G Open structure was tested to an average power of 445 W - where it operated at a 44.5MW with a 200 ns RF pulse at 50 Hz - without any sign that a limitation in the average power was being reached~\cite{halves2018}. For the nominal fill time, this would allow a repetition rate of 490~Hz at an input power of 19 MW which would need to be reduced to 450 Hz to be a harmonic of the 50 Hz mains. It is expected that the repetition rate will be able to exceed this. PSI's TW gun has been proposed to operate with an average power of 2.64 kW~\cite{PSI_TWgun2017} which would equal a repetition rate of 2.9 kHz. Further thermal analysis will be performed to understand the average power, and therefore repetition rate, limitations of the gun.

\begin{table}[]
    \centering
    \resizebox{0.9 \columnwidth}{!}{
    \begin{tabular}{||c c c||} 
     \hline
     Parameter & Value & Unit\\ [0.5ex] 
     \hline\hline
     Length & 216 & mm \\
     Regular Cells & 24 & \\
     Phase Advance& 120 & degs \\
     Frequency & 11.994 & GHz \\
     Attenuation & -2.23 & dB \\
     Power & 19 & MW \\       
     Gradient  & 65 & MV/m \\       
     Peak Cathode field & 91 & MV/m \\ 
     Fill time & 48 & ns \\
     Repetition Rate & 450 & Hz \\
     \hline\hline
     
    \end{tabular}
    }
    \caption{Summary of the travelling wave gun properties.}
    \label{tab:RFparameters}
\end{table}

\section{Power System}

\begin{figure}
    \centering
    \includegraphics[width=0.9\linewidth]{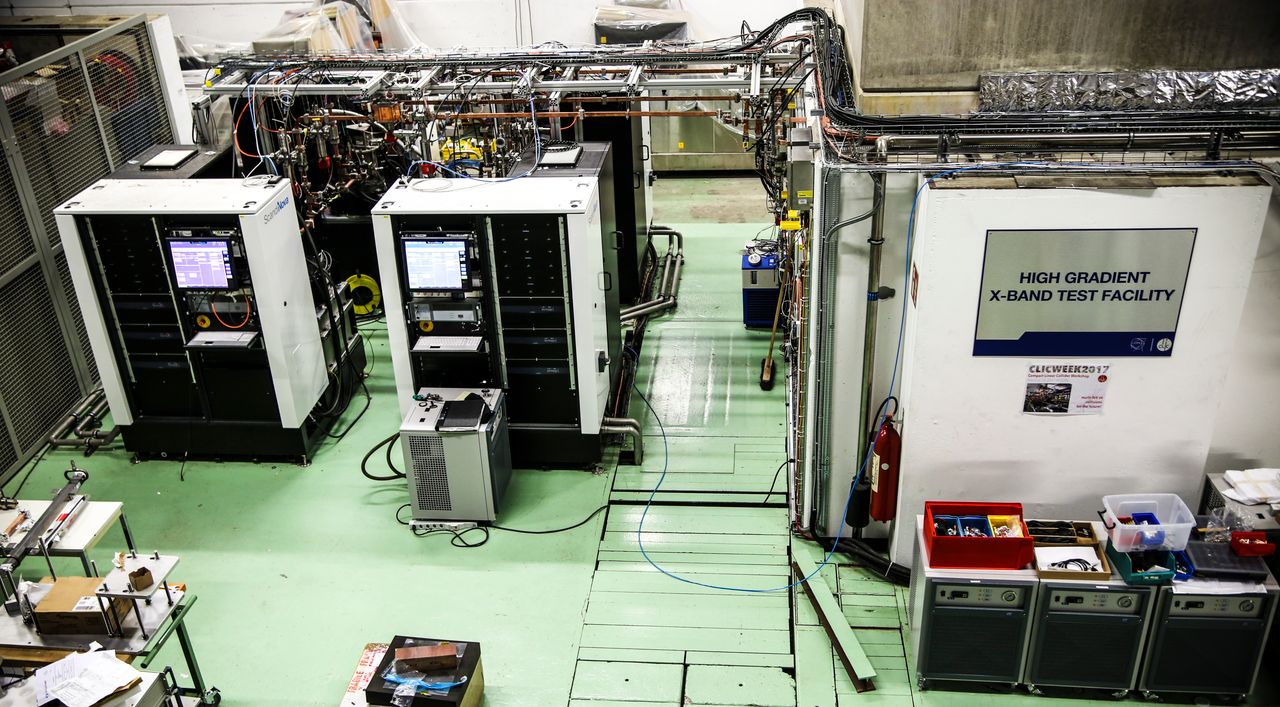}
    \caption{Xbox~3 test stand at CERN which uses 400 Hz Canon E37113 6 MW klystron.}
    \label{fig:Xbox3}
\end{figure}

\begin{figure}
    \centering
    \includegraphics[width=0.9\linewidth]{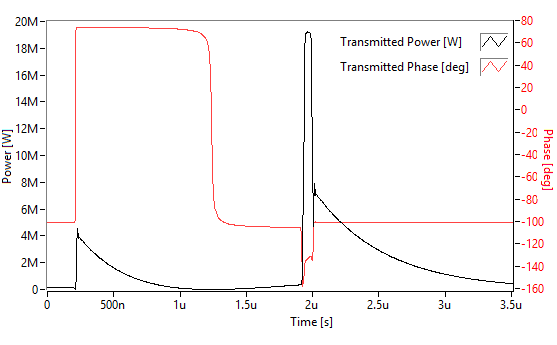}
    \caption{A simulation of the Xbox~3 pulse compressor outputting a 19 MW pulse from a 6 MW, 2$\mu$s input pulse.}
    \label{fig:SLED}
\end{figure}

Whether this repetition rates could be achieved is dependent on the RF power system. The current design is optimised for an input RF power of 19~MW, giving a gradient of 65 MV/m. The design could be powered by a Canon E37113 6 MW klystron feeding a pulse compressor operating at a gain of 3.2~\cite{Volpi}. Such klystrons are designed to operate at 400 Hz for a 5~$\mu$s pulse and proposed to operate at 1~kHz with a 2~$\mu$s RF pulse. A simulation of the pulse compressor used in CERN's Xbox~3, displayed in Figure~\ref{fig:SLED}, illustrates that this gain is achievable for a $2\mu$s input RF pulse. Furthermore, CPI and Canon have research programmes to develop high repetition rate klystrons operating at 10 MW. Using these it may become possible to generate the power through the weaving of two klystron pulses as is also at CERN's third X-band Test Stand (Figure~\ref{fig:Xbox3})~\cite{T.Lucas18}.

\begin{table}[]
    \centering
    \resizebox{ 0.8 \columnwidth}{!}{
    \begin{tabular}{||c c c||} 
     \hline
     Bunch Charge [pC] & 40  & 80  \\ 
     \hline\hline
     Laser spot max. radius [um]  & 282 & 400  \\
     norm. intrinsic emittance [mm mrad/mm] & 0.55  & 0.55  \\  
     Laser size at 250mm [um]  & 291 & 403  \\
     Laser pulse length RMS [fs] & 100 & 100 \\
     Bunch length RMS [fs] & 510 & 550  \\
     norm. emittance [mm mrad] & 0.32  & 0.6  \\     
     Beam Energy [MeV]  & 14.14 & 14.14  \\
     Beam Energy Spread [$\%$]   & 0.09 & 0.14 \\
     
     \hline\hline
     
    \end{tabular}
    }
    \caption{Summary of the travelling wave gun's laser and electron beam properties.}
    \label{tab:Laser_and_beam}
\end{table}

\begin{figure*}
    \centering
    \includegraphics[width=0.8\linewidth]{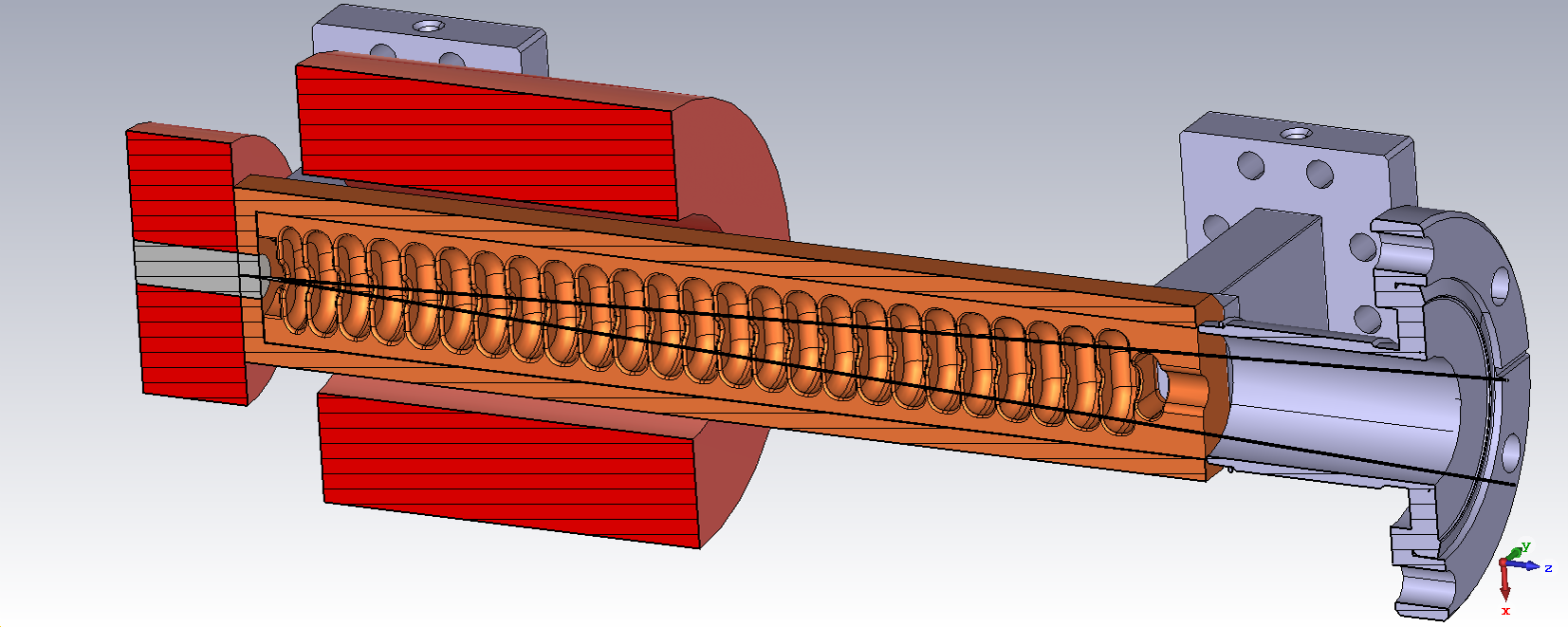}
    \caption{One symmetric half of the RF design (vacuum) of the travelling wave RF photogun with the notable features labelled.}
    \label{fig:structuredesign}
\end{figure*}

\section{Laser coupling}

A possible criticism of this style of RF photogun is the acceptance angle of the accelerator for a laser. If one were to modify a typical CLIC-G design, which has no gap, into a TW RF photogun an acceptance angle of 0.6 degrees would be possible if the laser were to be coupled through the irises. Using such a method may make it difficult to couple in the laser without obstructing outgoing electrons, given the compactness of the X-band structure's geometry. The addition of the gap has opened up this angle quite significantly to an angle up to 3.6 degrees. This angle could be increased greatly by placing a window on the side of the gun and coupling through the side of the structure, again using the gap. A limitation of the coupling is the width of the gap. The beam must be carefully thread through the 1~mm gap with minimal interception of the wall. This ultimately limits the largest spot size one can achieve. Table~\ref{tab:Laser_and_beam} illustrates the laser radius on the cathode and the radius of the laser at 250~mm, which is approximately the length from the cathode to the end of the structure. Calculating the Rayleigh Range and the beam envelope, it is found that the greatest bunch charge achievable is approximately 120~pC. This could be increased through an increase to the charge density on the cathode but may lead to an increased emittance.

\begin{figure}
    \centering
    \includegraphics[width=0.8\linewidth]{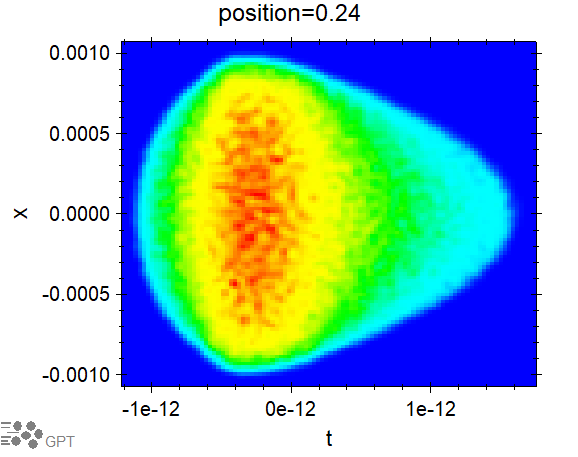}
    \caption{The longitudinal distribution of a 80 pC bunch}
    \label{fig:x_vs_t_distribution}
\end{figure}

\begin{figure}
    \centering
    \includegraphics[width=0.8\linewidth]{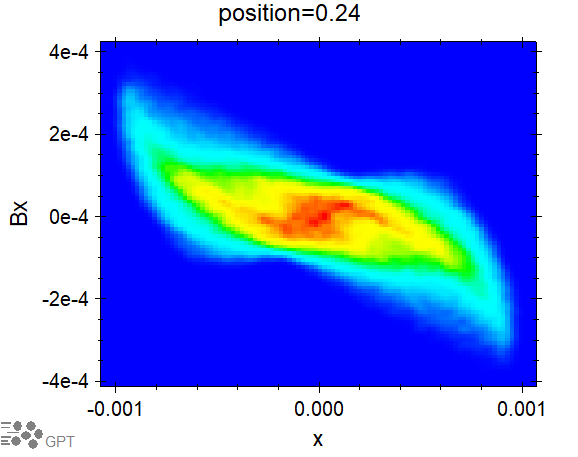}
    \caption{The phase space distribution of a 80 pC bunch}
    \label{fig:Bx_vs_x_distribution}
\end{figure}

\section{Main Solenoid and Bucking Coil}

After the RF design was finalised, it was incorporated into GPT and a solenoid was added to counter the defocusing effect at the end of the structure, to keep the beam from colliding with the cell irises and to maintain the low emittance in the structure. Along with the main solenoid, a bucking coil was added to reduce the field at the cathode to prevent an increase in the intrinsic emittance caused by local magnetic fields. GPT was used to optimise the length and field strength of the main solenoid as well as the bucking coil. The conditions for the optimisation were that the beam's divergence was less than 100 $\mu$rad, the field at the cathode was less than 20 $\mu$T, and the emittance was less than that without the solenoid. The axial magnetic field distribution is illustrated in Figure~\ref{fig:EandHfield} where the B~field at the cathode is observed to be approximately zero (6 $mu$T) and the focusing field of the main solenoid covers the first half of the structure.

\section{Beam Dynamics}
\begin{figure}
    \centering
    \includegraphics[width=\linewidth]{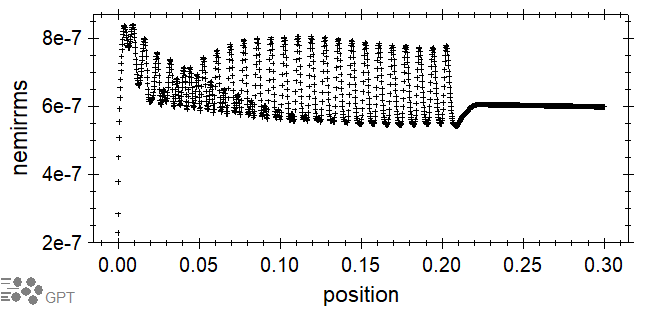}
    \caption{The evolution of the emittance over the gun.}
    \label{fig:emittance}
\end{figure}

With the optimised solenoids from GPT and the axial electric field distribution from the CST model, the entire gun was simulated from start-to-end in GPT. The off-axis electric fields were extrapolated from the imported on-axis field distribution to match that of a TM mode structure. These do not include the quadrupole component expected from a dual-fed input coupler which will be investigate in future work. The cathode was simulated with an intrinsic emittance of 0.55 mm mrad/mm as has been used in~\cite{PSI_TWgun2016} and the laser spot size on the cathode was set to 400$\mu$m for an 80~pC bunch. For different charges the laser spot size was scaled according to $r \propto \sqrt{Q}$ to keep a constant charge density. GPT's optimiser was used to find the appropriate RF power (between 15 and 20 MW), RF phase and laser pulse length (between 50~fs and 1~ps) to minimise the emittance and energy spread. For values between 50 and 500 fs the pulse length was found to have little effect on the beam quality. The spatial distribution and phase-space distributions of the bunch which resulted from these optimisations are illustrated in Figures~\ref{fig:x_vs_t_distribution} and~\ref{fig:Bx_vs_x_distribution}, respectively. Figure~\ref{fig:emittance} illustrates the emittance over the length of the gun. Table~\ref{tab:Laser_and_beam} summarises the results of the output electron beam quality for bunch charges of 40 and 80 pC.

\section{Conclusion and Future Steps}
A design for a travelling wave X-band photogun based on a CLIC prototype made from milled halves was studied. The unique gap in the geometry was used to couple in the laser to the cathode without obstructing the output electrons. With a high field solenoid and bucking coil, the start-to-end beamline simulations were performed in GPT illustrating that the gun could produce 80~pC electron bunches at 14.15~MeV with a low energy spread of 0.14~$\%$ energy spread at an emittance of 0.6~mm~mrad. The short fill time of 48 ns allows the gun to operate at very high repetition rates. There are various future steps for this gun. To finalise this design of this gun, the surface fields on the input coupler will be optimised and an investigation into the effects of the quadrupole component at the cathode will be studied. Separate to this, a high power version of this gun will be developed in the aim of producing a lower emittance gun which can compete with the current stare-of-the-art standing wave guns.

\section{Acknowledgements}
The authors would sincerely like to thank Xavier Stragier, Marco van der Sluis and Harry Doorn from Eindhoven University of Technology, and Walter Wuensch and Nuria Catalan Lasheras from CERN for their input and discussions about the project.

\newpage

\bibliographystyle{elsarticle-num}
\bibliography{References}

\end{document}